\documentclass[%
superscriptaddress,
preprintnumbers,
tightenlines,
showpacs,
showkeys,
nofootinbib,
amsmath,
amssymb,
aps,
prd,
floatfix,
]{revtex4-1}

\usepackage{graphicx}
\usepackage{color}

\newcommand{\rmE}{{\rm E}}

\newcommand{\dof}{{\rm d.o.f.}}

\begin{document}
\title{Critical endline of the finite temperature phase transition for 2+1 flavor QCD around the SU(3)-flavor symmetric point}

\author{Yoshinobu Kuramashi}
\affiliation{Faculty of Pure and Applied Sciences, University of Tsukuba, Tsukuba, Ibaraki 305-8571, Japan}
\affiliation{Center for Computational Sciences, University of Tsukuba, Tsukuba, Ibaraki 305-8577, Japan}
\affiliation{RIKEN Advanced Institute for Computational Science, Kobe, Hyogo 650-0047, Japan}

\author{Yoshifumi~Nakamura}\email[]{nakamura@riken.jp}
\affiliation{RIKEN Advanced Institute for Computational Science, Kobe, Hyogo 650-0047, Japan}
\affiliation{Graduate School of System Informatics, Department of Computational Sciences, Kobe University, Kobe, Hyogo 657-8501, Japan}

\author{Shinji Takeda}\email[]{takeda@hep.s.kanazawa-u.ac.jp}
\affiliation{Institute of Physics, Kanazawa University, Kanazawa 920-1192, Japan}
\affiliation{RIKEN Advanced Institute for Computational Science, Kobe, Hyogo 650-0047, Japan}

\author{Akira Ukawa}
\affiliation{RIKEN Advanced Institute for Computational Science, Kobe, Hyogo 650-0047, Japan}

\date{\today}
\begin{abstract}
We investigate the critical endline of the finite temperature phase transition of  QCD around the SU(3)-flavor symmetric point at zero chemical potential.
We employ the renormalization-group improved Iwasaki gauge action and non-perturbatively $O(a)$-improved Wilson-clover  fermion action.
The critical endline is determined by using the intersection point of kurtosis, employing the multi-parameter, multi-ensemble reweighting method to calculate observables off the SU(3)-symmetric point, at the temporal size $N_{\rm T}$=6  and lattice spacing as low as $a \approx 0.19$ fm.
We confirm that the slope of the critical endline takes the value of $-2$,  and find that the  second derivative is positive, at the SU(3)-flavor symmetric point on the Columbia plot parametrized with the strange quark mass $m_s$ and degenerated up-down quark mass $m_{\rm l}$. 
\end{abstract}


\preprint{UTHEP-687, UTCCS-P-84, KANAZAWA-16-05}

\maketitle
\section{Introduction}\label{sec:int}

The Columbia phase diagram plot~\cite{ColumbiaPlot90} provides a visualization of the finite-temperature phases  of 2+1 flavor QCD at zero chemical potential  in the plane of the light quark mass and strange quark mass $( m_{\rm l}$, $m_{\rm s}$). 
In the small quark mass region, it is believed that the transition is of first order~\cite{PisarskiWilczek84}, which turns into a region of crossover as quark masses are increased.  
The boundary which separates the two regions is the critical endline (CEL) which belongs to the Z$_2$ universality class~\cite{GavinGokschPisarski94}. 

There is a longstanding issue that the results for critical endpoint (CEP) at the SU(3)-flavor symmetric point ($m^{\rm sym}$) obtained 
by Wilson type and staggered type fermion actions are inconsistent~\cite{Iwasaki96,Kaya99,Karsch01,Philipsen07,Smith11,Endrodi07,Ding11}.
Recently, we have investigated CEP with degenerate $N_{\rm f}=3$ dynamical flavors of non-perturbatively $O(a)$-improved Wilson fermion action, 
and determined its location by the intersection points of kurtosis for the temporal sizes $N_{\rm T}=4$, $6$ and $8$~\cite{cep3ft}.  The continuum extrapolation implies a non-zero value $m_{\rm PS, CEP}\approx 300$~MeV for the pseudoscalar meson mass.  Scaling violations are large, however, necessitating further studies at larger $N_{\rm T}$  to obtain conclusive results for CEP. 

In this article, we explore the properties of CEL away from $m^{\rm sym}$.  In particular we ask how CEL curve away from $m^{\rm sym}$.
This is a first step to obtain a comprehensive view on the relation of CEL and the physical point for which the strange quark mass is significantly heavier than the degenerate up-down quark mass. 
To set the stage for our analysis, let us consider the kurtosis $K_{\cal O}$ of some quantity ${\cal O}$ which can be either a gluonic or quark quantity.  The kurtosis generally depends on the quark masses $m_{\rm u}, m_{\rm d}, m_{\rm s}$, and its Taylor expansion around $m^{\rm sym}$ will have a form $K_{\cal O}=K_0 + (\delta m_{\rm u}+ \delta m_{\rm d}+ \delta m_{\rm s}) K_1 + O(\delta m^2)$ where $\delta m_q (q={\rm u}, {\rm d}, {\rm s})$ denotes the difference from the SU(3)-symmetric value $m^{\rm sym}$.   Therefore, if one varies  the quark masses while keeping the average over the three quark masses, {\it i.e.}, 
$\delta m_{\rm u}+ \delta m_{\rm d}+ \delta m_{\rm s}=0$, the kurtosis remains unchanged up to second order in the variation of the quark masses.  
For degenerate up and down quark mass, we have $\delta m_{\rm l} = \delta m_{\rm u}=\delta m_{\rm d}$, and hence the change becomes
\begin{equation}
\label{deltam}
\delta m_{\rm s}= -2 \delta m_{\rm l} .
\end{equation}
This means that the slope of CEL at $m^{\rm sym}$ should take the value $-2$ on the Columbia plot.

There are no such constraints on the second derivative of CEL with respect to $m_{\rm l}$ at $m^{\rm sym}$.  
If it is positive, CEL  would smoothly curve  up to the tricritical point  $m_{\rm s} = m_{\rm s}^{\rm tric}$ located on the axis for the strange quark mass around which CEL is expected to behave as 
$m_{\rm s} - m_{\rm s}^{\rm tric} \sim  m_{\rm l}^{2/5}$~\cite{Rajagopal95}. 
So far, a lattice QCD result obtained by using staggered fermions at $N_{\rm T}=4$ with a lattice spacing $a \approx 0.3$ fm supports such a curve~\cite{Philipsen07}.

%
This paper is organized as follows.
In Section~\ref{sec:sim} we present the simulation details including the parameters and the simulation algorithm. 
Our numerical results are presented in Section~\ref{sec:res}.  In Section~\ref{sec:sum}, we provide a brief conclusion.

\section{Simulation details}\label{sec:sim}

To determine CEL away from CEP, 
we perform kurtosis intersection analysis using a multi-parameter, multi-ensemble reweighting method.
The details of this method is described in Refs.~\cite{cep3ft,cep3fd}.

Calculations are made at a temporal lattice size $N_{\rm T}=6$ and the spacial sizes $N_{\rm L}=10$, $12$, $16$ and $24$
with $N_{\rm f}=3$ degenerate flavors of dynamical quarks 
using the Iwasaki gluon action~\cite{iwasaki} and the non-perturbatively $O(a)$-improved Wilson fermion action~\cite{csw}.
All observables for $N_{\rm f}=2+1$ QCD are computed by using a $\kappa$ reweighting.
The periodic boundary condition is imposed for gluon fields while the anti-periodic boundary condition is employed for quark fields.
We use a highly optimized HMC code~\cite{BQCD}, applying mass preconditioning~\cite{mprec} and 
RHMC~\cite{RHMC}, 2nd order minimum norm integration scheme~\cite{Omelyan},
putting the pseudo fermion action on multiple time scales~\cite{mtime}
and a minimum residual chronological method~\cite{chronological} to choose the starting guess for the solver.
We generate $63$ ensembles of $O(100,000)$ trajectories  on HA-PACS and COMA at University of Tsukuba,
System E at Kyoto University and PRIMERGY CX400 tatara at Kyushu University.  
Measurements are done at every 10th trajectory and statistical errors are estimated by the jackknife method with the bin size of $O(1,000)$ configurations.
In Table~\ref{tab:ensemble}, we summarize the simulation parameters and statistics.

\begin{table}[!hbt]
\caption{Simulation parameters and statistics.}
\label{tab:ensemble}%
\begin{ruledtabular}
\begin{tabular}{ll|r|r|r|r}
               &                &  \multicolumn{4}{c}{\# of conf.} \\
$ \beta $ & $\kappa$ & $N_{\rm L}=10$ & $N_{\rm L}=12$ & $N_{\rm L}=16$ & $N_{\rm L}=24$ \\
\hline
$ 1.715 $&$ 0.140900 $&$ 7300 $&$ 4200 $&$      $&      \\
$ 1.715 $&$ 0.140920 $&$ 8000 $&$ 3800 $&$      $&      \\
$ 1.715 $&$ 0.140940 $&$ 8000 $&$ 8600 $&$ 7900 $&      \\
$ 1.715 $&$ 0.140950 $&$ 8100 $&$ 7900 $&$ 8000 $&      \\
$ 1.715 $&$ 0.140960 $&$ 9500 $&$ 8800 $&$ 7900 $&      \\
$ 1.715 $&$ 0.140970 $&$ 9400 $&$ 6900 $&$ 8400 $&      \\
$ 1.715 $&$ 0.140980 $&$ 9699 $&$ 6800 $&$ 8800 $&      \\
$ 1.715 $&$ 0.140990 $&$ 9500 $&$ 6300 $&$ 8599 $&      \\
$ 1.715 $&$ 0.141000 $&$10000 $&$ 8700 $&$ 9400 $&      \\
$ 1.715 $&$ 0.141010 $&$ 8800 $&$ 8500 $&$ 8300 $&      \\
$ 1.715 $&$ 0.141020 $&$ 8700 $&$ 8900 $&$ 8600 $&      \\
$ 1.715 $&$ 0.141100 $&$ 8600 $&        &        &      \\ \hline
$ 1.73  $&$ 0.140420 $&$ 7900 $&$ 7900 $&$ 8900 $&$5250$\\
$ 1.73  $&$ 0.140430 $&$ 7900 $&$ 7900 $&$ 8900 $&$5100$\\
$ 1.73  $&$ 0.140440 $&$ 7900 $&$ 7900 $&$ 8900 $&$5200$\\
$ 1.73  $&$ 0.140450 $&$ 8600 $&$ 7900 $&$ 7300 $&$4900$\\ \hline
$ 1.75  $&$ 0.139620 $&$12200 $&$10450 $&$10100 $&\\
$ 1.75  $&$ 0.139640 $&$12900 $&$10400 $&$10100 $&\\
$ 1.75  $&$ 0.139660 $&$11700 $&$10450 $&$10100 $&\\
$ 1.75  $&$ 0.139680 $&$12900 $&$10400 $&$ 9700 $&\\
$ 1.75  $&$ 0.139700 $&$ 8378 $&$ 4900 $&$10100 $&\\
\end{tabular}
\end{ruledtabular}
\end{table}

In this study, we use the susceptibility, $\chi$, of the quark condensate to determine the transition point, 
and its kurtosis, $K$, for intersection analysis to locate CEL.
The quark condensate, $\Sigma$, and skewness, $S$, are used to check that the transition point is determined appropriately.
The quantities $\Sigma$, $\chi$, $S$ and $K$ are defined by
\begin{eqnarray}
\begin{split}
\label{eq:vk}
\Sigma=&\frac{\langle Q_1\rangle}{N_{\rm L}^3 N_{\rm T}},\label{eqn:exp}\\
\chi  =&\frac{\langle Q_2\rangle - \langle Q_1\rangle^2}{N_{\rm L}^3 N_{\rm T}},\label{eqn:sus}\\
S  =&\frac{\langle Q_3\rangle - 3\langle Q_2\rangle \langle Q_1\rangle+2\langle Q_1\rangle^3}{\left(\langle Q_2\rangle - \langle Q_1\rangle^2\right)^{3/2}},\\
K  =&\frac{\langle Q_4\rangle - 4\langle Q_3\rangle \langle Q_1\rangle -3\langle Q_2\rangle^2+12\langle Q_2\rangle \langle Q_1\rangle^2-6\langle Q_1\rangle^4}{\left(\langle Q_2\rangle - \langle Q_1\rangle^2\right)^{2}},
\end{split}
\end{eqnarray}
with
\begin{eqnarray}
\begin{split}
Q_1=&N_{\rm f}{\rm tr}[D_f^{-1}]\,,\\
Q_2=&-N_{\rm f}{\rm tr}[D_f^{-2}]+(N_{\rm f}{\rm tr}[D_f^{-1}])^2\,,\\
Q_3=&2N_{\rm f}{\rm tr}[D_f^{-3}]-3N_{\rm f}^2{\rm tr}[D_f^{-2}]{\rm tr}[D_f^{-1}]+(N_{\rm f}{\rm tr}[D_f^{-1}])^3\,,\\
Q_4=&-6N_{\rm f}{\rm tr}[D_f^{-4}]+8N_{\rm f}^2{\rm tr}[D_f^{-3}]{\rm tr}[D_f^{-1}]+3(N_{\rm f}{\rm tr}[D_f^{-2}])^2-6N_{\rm f}{\rm tr}[D_f^{-2}](N_{\rm f}{\rm tr}[D_f^{-1}])^2+(N_{\rm f}{\rm tr}[D_f^{-1}])^4\,,\\
\end{split}
\end{eqnarray}
where 
\begin{eqnarray}
\begin{split}
D_f=&\frac{1}{2\kappa_{f}}+\frac{i}{4}c_{\rm sw} \sigma_{\mu\mu} F_{\mu\nu}(n)\delta_{m,n}
-\frac{1}{2}\sum_{\mu=1}^{4}\left[(1-\gamma_\mu)U_{\mu}(n)\delta_{n,m+\hat\mu} + (1+\gamma_\mu)U_{-\mu}(n)\delta_{n,m-\hat\mu} \right]\,.\\
\end{split}
\end{eqnarray}
There are a few choices for the quark condensate in $N_{\rm f}=2+1$ QCD.
For example, eq.~(\ref{eq:vk}) for $\bar ss$ is defined by $\kappa_f=\kappa_{\rm s}$ and $N_{\rm f}=1$.
In this study, we choose $\kappa_f=\kappa_{\rm l}$ and $N_{\rm f}=2$
since the signal of for $\bar uu+\bar dd$ and its higher moments turns out to be better than the others.
Even if we made different choices, we expect to obtain the same results
because all of such ``order parameters'' would behave equally as pure magnetization in the thermodynamic limit.

\section{Results}\label{sec:res}

\begin{figure}[!hbt]
  \centering
  \includegraphics[bb=0 0 340 255,width=0.4\textwidth]{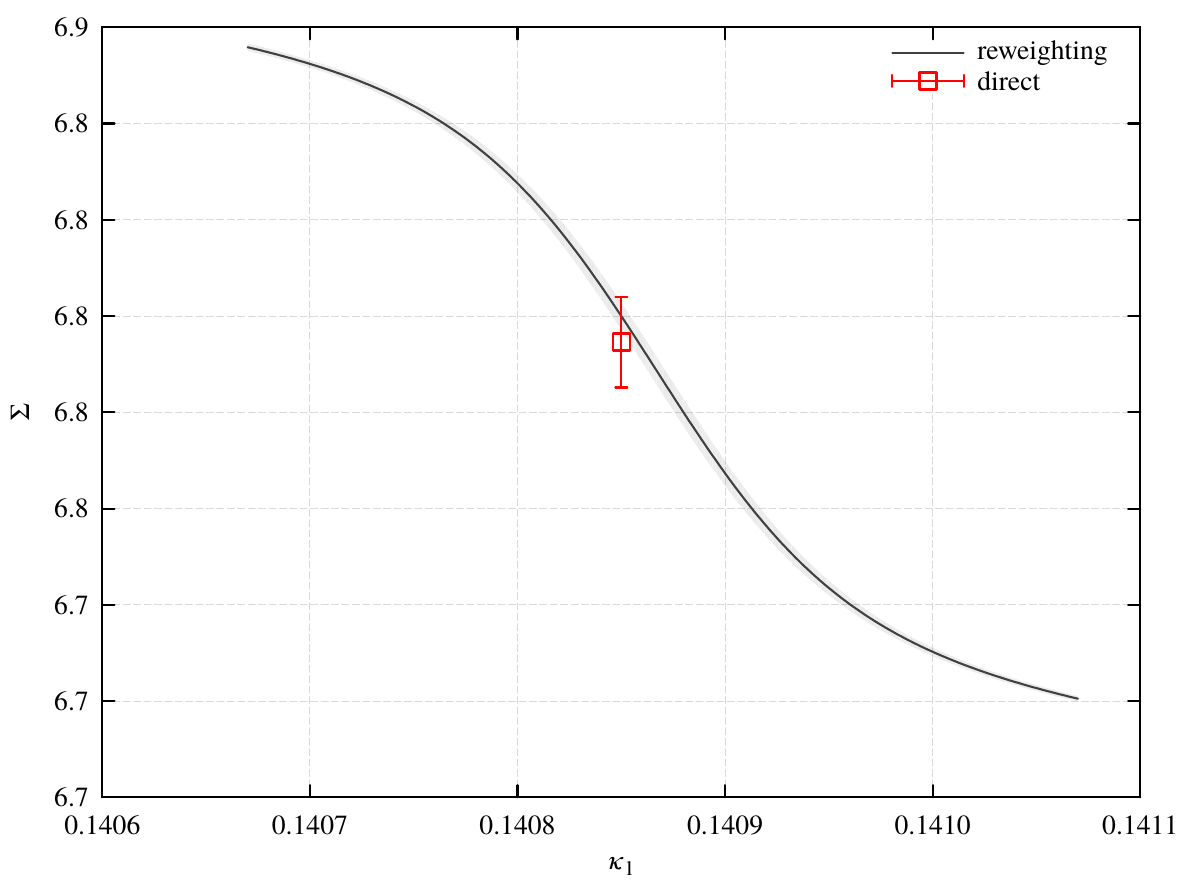}
  \includegraphics[bb=0 0 340 255,width=0.4\textwidth]{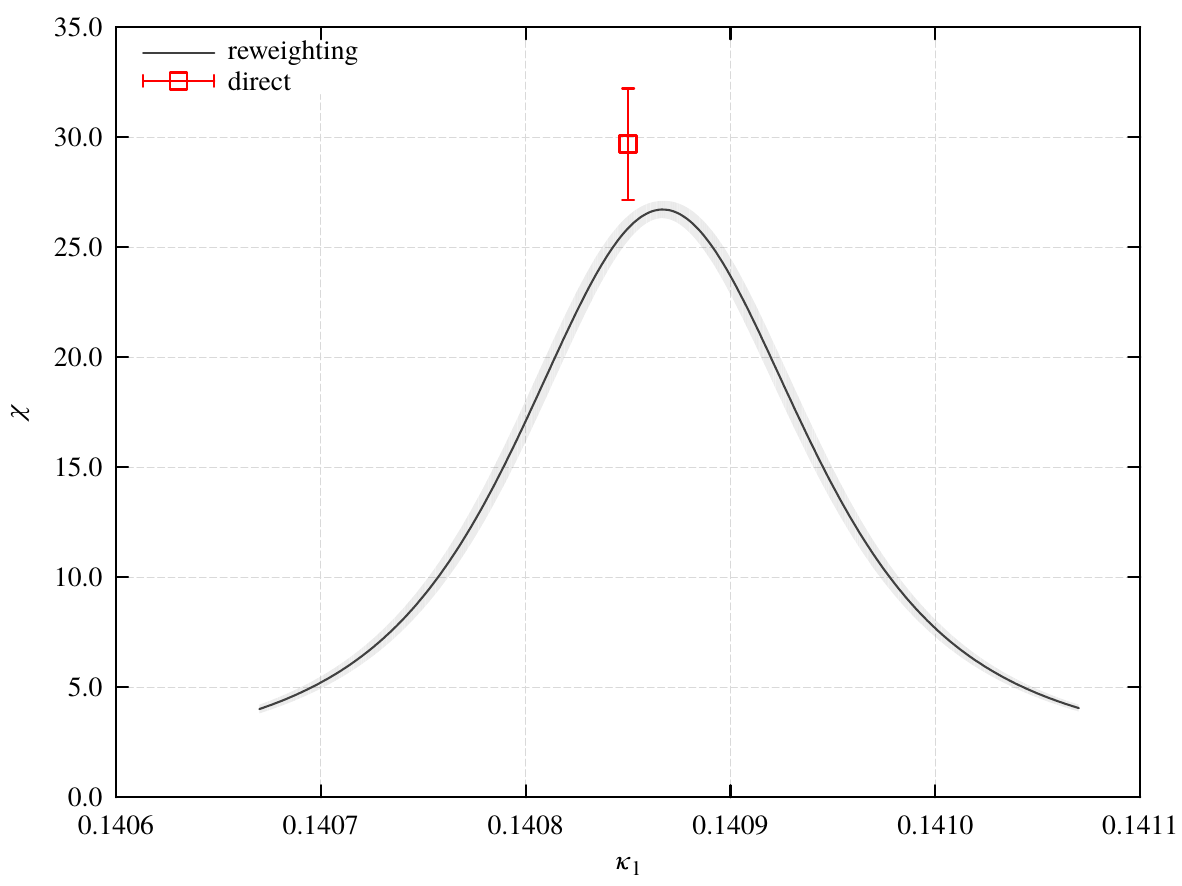}\\
  \includegraphics[bb=0 0 340 255,width=0.4\textwidth]{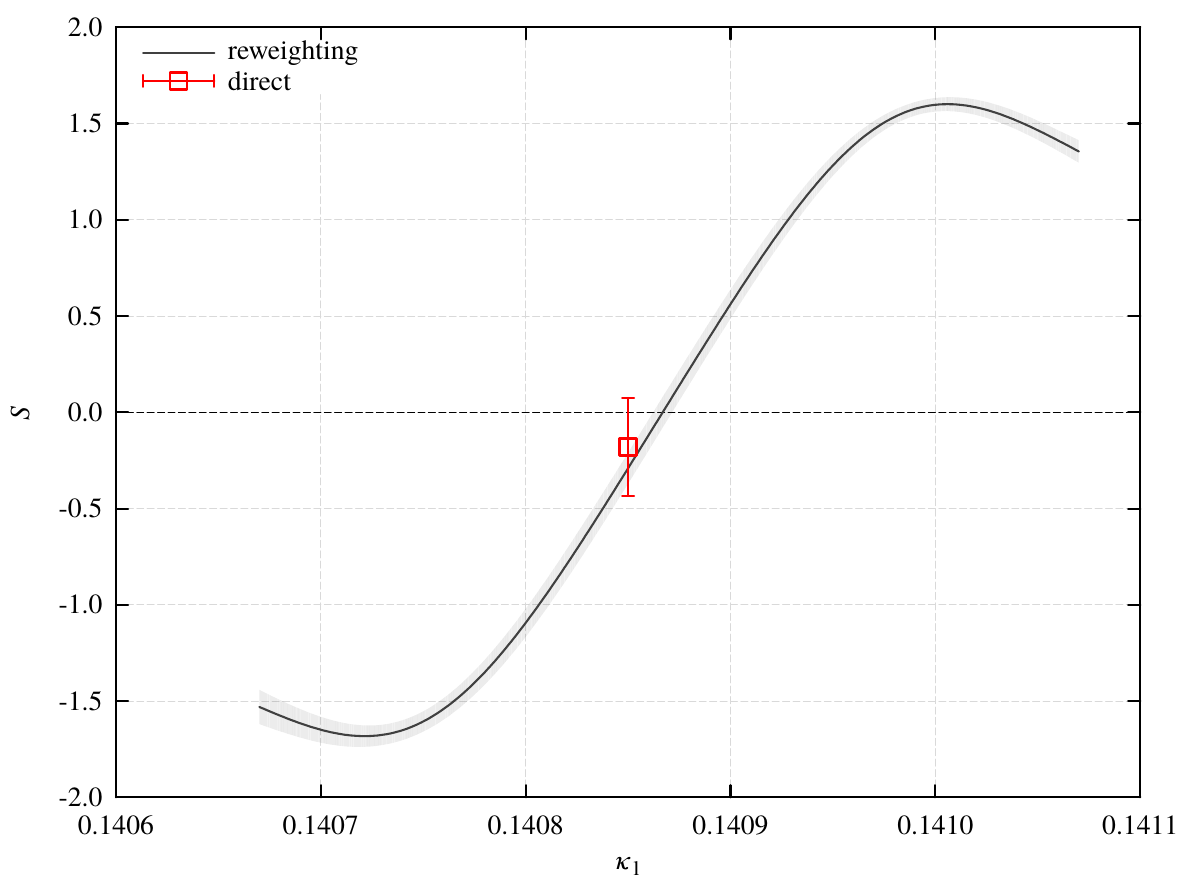}
  \includegraphics[bb=0 0 340 255,width=0.4\textwidth]{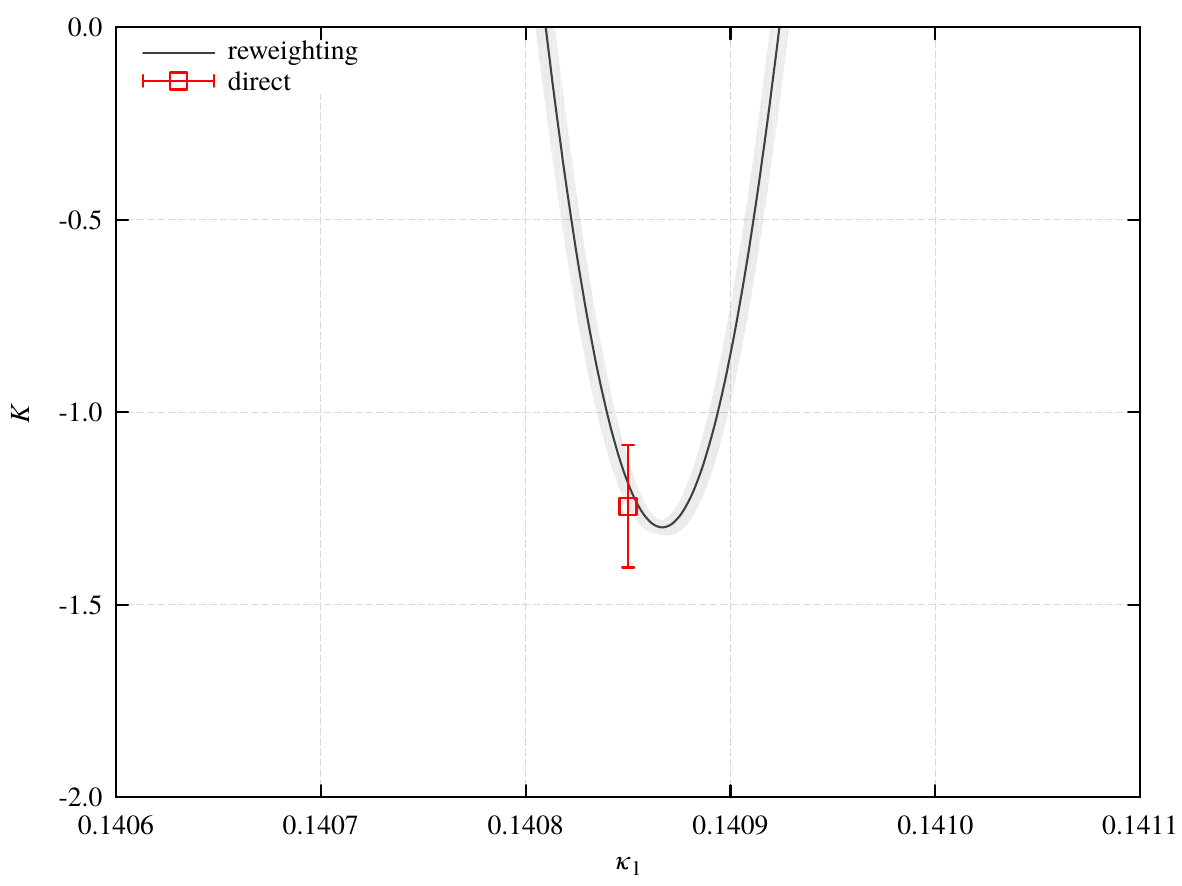}
  \caption{$\Sigma$(left up), $\chi$(right up), $S$(left bottom) and $K$(right bottom) as a function of $\kappa_{\rm l}$ obtained by reweighting of data in Table~\ref{tab:ensemble} (curves)  and those at $\beta=1.73$,
  $\kappa_{\rm s}=0.13950$ $N_{\rm L}=12$ obtained by the direct simulation at $\kappa_{\rm l}=0.140850$ (square data points). }
  \label{fig:reweighting}
\end{figure}

We check first the validity of the reweighting from $N_{\rm f}=3$ to $N_{\rm f}=2+1$.
Figure~\ref{fig:reweighting} compares $\Sigma$, $\chi$, $S$ and $K$ obtained by the reweighting of the $N_{\rm f}=3$ runs at $\beta=1.73$ with a direct $N_{\rm f}=2+1$simulation 
at $(\beta,N_{\rm L},\kappa_{\rm l},\kappa_{\rm s})=(1.73,12,0.140850,139500)$ as a function of $\kappa_{\rm l}$. 
They are in good agreement with each other. 

\begin{figure}[!hbt]
  \centering
  \includegraphics[bb=0 0 794 198,width=1.\textwidth]{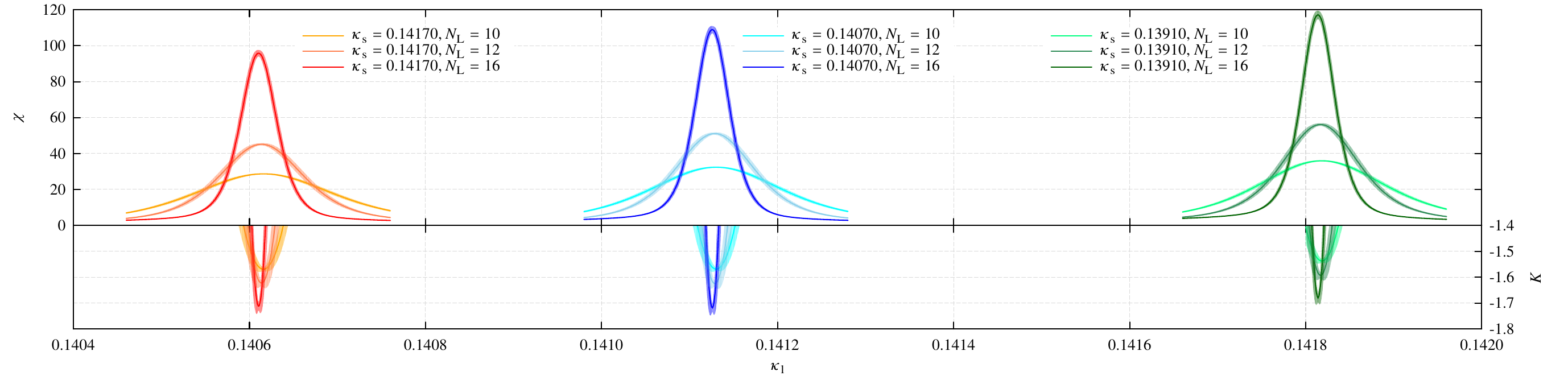}
  \includegraphics[bb=0 0 794 198,width=1.\textwidth]{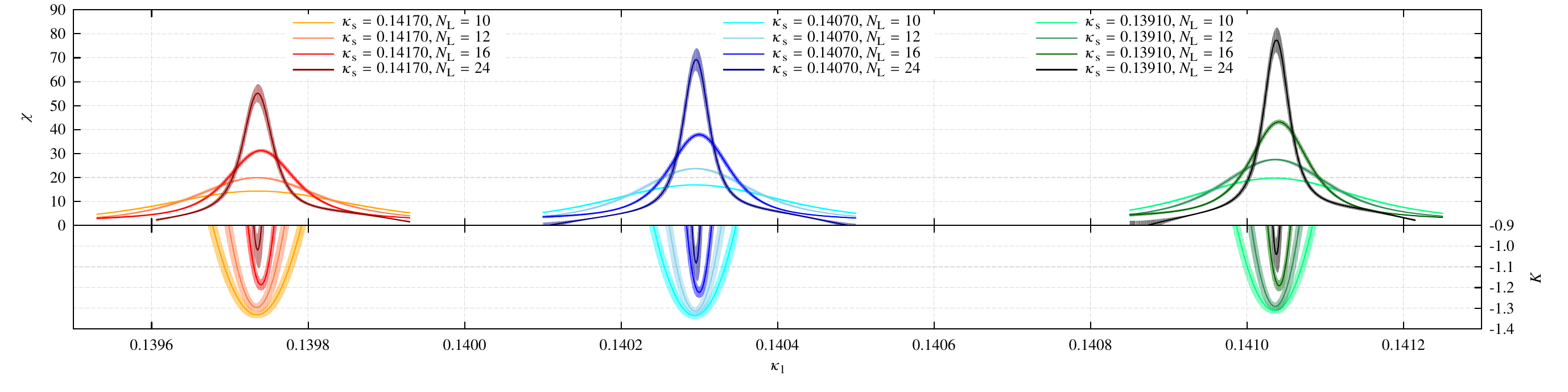}
  \caption{$\chi$ and $K$ for $\kappa_{\rm s}=0.13910, 0.14070, 0.14170$ as a function of $\kappa_{\rm l}$ at $\beta=1.715$(upper) and $1.73$(lower).}
  \label{fig:typical_sus_krt}
\end{figure}

The reweighting results for the susceptibility and kurtosis are illustrated in Fig.~\ref{fig:typical_sus_krt} for $\beta=1.715$ and $1.73$  and three values of $\kappa_{\rm s}$ as functions of $\kappa_{\rm l}$.    
The peak position of susceptibility gives us a very precise value of $\kappa_{\rm l}$ for the transition point,  and a corresponding value of kurtosis, for each $\beta$, $\kappa_{\rm s}$ and $N_{\rm L}$.

\begin{figure}[!hbt]
  \centering
  \includegraphics[bb=0 0 340 255,width=.5\textwidth]{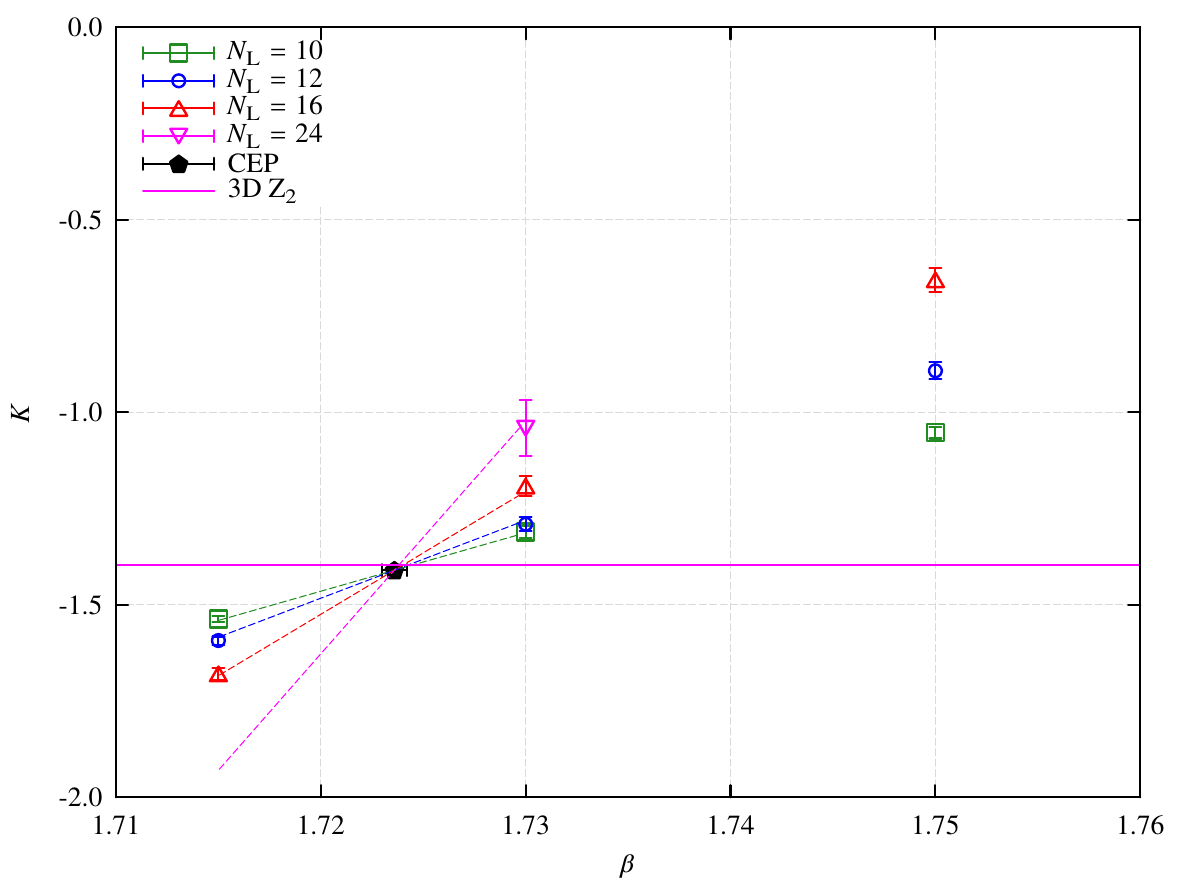}
  \includegraphics[bb=0 0 340 255,width=.5\textwidth]{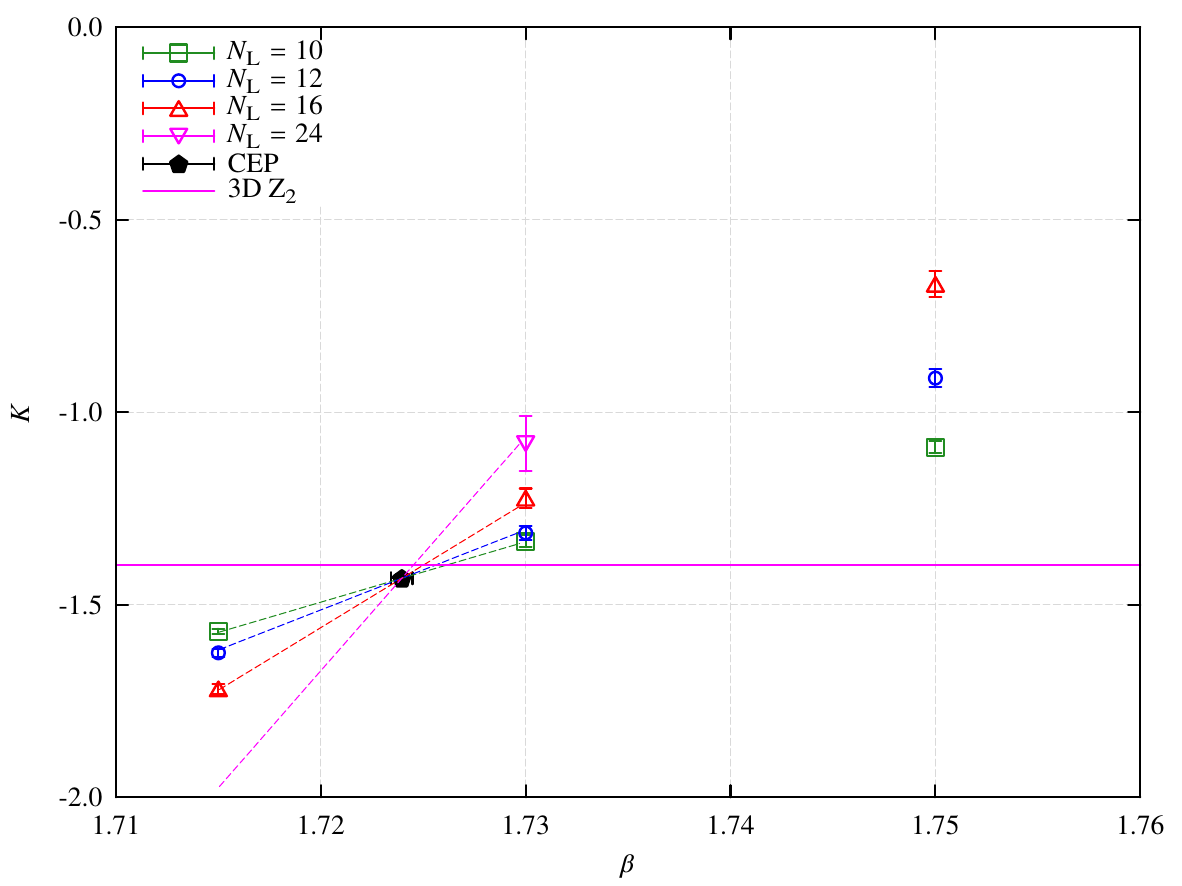}
  \includegraphics[bb=0 0 340 255,width=.5\textwidth]{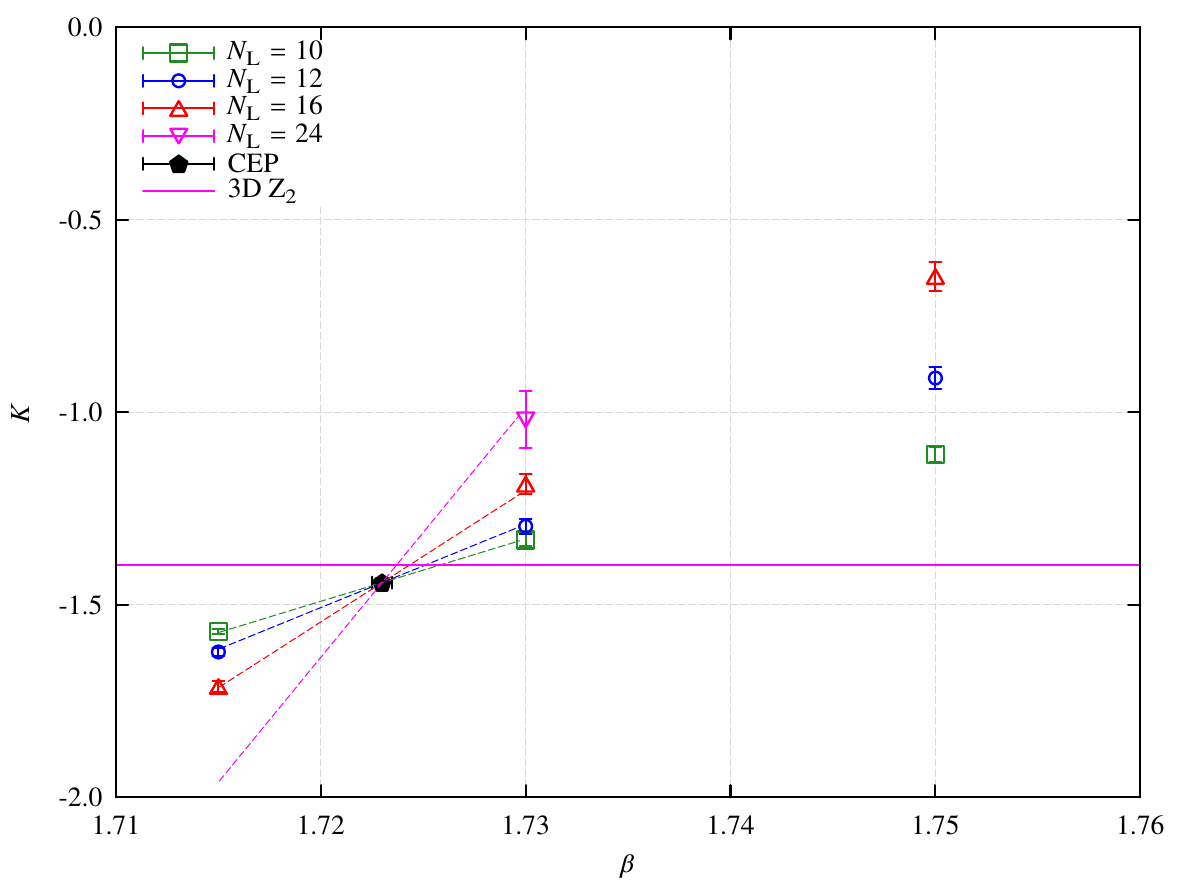}
  \caption{Typical kurtosis intersection plots. These are for $\kappa_{\rm s}=0.13910$(top), $0.14070$(middle) and $0.14170$(bottom).}
  \label{fig:typical_intersection}
\end{figure}

Figure~\ref{fig:typical_intersection} shows the kurtosis at the transition point for three values of $\kappa_{\rm s}$  in Fig.~\ref{fig:typical_sus_krt}.    We fit the kurtosis at $\beta=1.715$ and $1.73$ with a finite-size scaling ansatz given by 
\begin{equation}
\label{eq:intersection}
K=K_\rmE + AN_{\rm L}^{1/\nu}(\beta-\beta_\rmE)\,,
\end{equation}
where $\beta_\rmE$ and $K_\rmE$ are the values of $\beta$ and $K$ at each CEP and $\nu$ is the critical exponent along CEL.
The fitting results are summarized in Table~\ref{tab:beke}.
We find that $\nu$ and $K_\rmE$ are consistent with the values of the three-dimensional Z$_2$ universality class.
The results for $\kappa_{\rm s}<0.13910$ or $\kappa_{\rm s}>0.14170$ are too noisy to determine CEP  
because they are too far away from the original simulation points along the SU(3)-symmetric line.

Table~\ref{tab:beke} also lists the value of $\kappa_{\rm l, E}$.  To obtain them, we  go back to the analysis in Fig.~\ref{fig:typical_sus_krt}.  We extrapolate  $\kappa_{\rm l}$ for the peak position linearly  to the thermodynamic limit by $1/N_{\rm L}^3$ for each $\beta$ and $\kappa_{\rm s}$.  For given $\kappa_{\rm s}$, we then fit the results as a quadratic function of $\beta$, and calculate $\kappa_{\rm l, E}$ by setting $\beta=\beta_{\rm E}$.

\begin{table}[!hbt]
\caption{Fitting results for eq.~(\ref{eq:intersection}).
$\kappa_{\rm l,E}$ is obtained by interpolation to $\beta_\rmE$ after fitting with quadratic function of $\beta$ (see text for more details).
$\kappa_{\rm s,E}$ has no error because we fix them as target parameters of the reweighing method.
}
\label{tab:beke}
\begin{ruledtabular}
\begin{tabular}{lllllll}
$\kappa_{\rm s,E}$&$\beta_{\rm E}$&$\kappa_{\rm l,E}$&$K_{\rm E}$&$\nu$&$A$&$\chi^2/\dof$\\
\colrule
$0.13910$&$1.72359 ( 61 )$&$0.1413751 ( 20 )$&$-1.410 ( 13 )$&$0.633 ( 46 )$&$0.40 ( 12 )$&$0.58$\\
$0.13920$&$1.72364 ( 60 )$&$0.1413321 ( 21 )$&$-1.411 ( 13 )$&$0.635 ( 46 )$&$0.40 ( 12 )$&$0.57$\\
$0.13930$&$1.72370 ( 60 )$&$0.1412884 ( 21 )$&$-1.412 ( 13 )$&$0.637 ( 45 )$&$0.41 ( 12 )$&$0.55$\\
$0.13940$&$1.72376 ( 59 )$&$0.1412441 ( 21 )$&$-1.414 ( 12 )$&$0.638 ( 45 )$&$0.41 ( 11 )$&$0.54$\\
$0.13950$&$1.72381 ( 58 )$&$0.1411993 ( 20 )$&$-1.415 ( 12 )$&$0.639 ( 44 )$&$0.42 ( 11 )$&$0.53$\\
$0.13960$&$1.72386 ( 57 )$&$0.1411540 ( 21 )$&$-1.416 ( 12 )$&$0.641 ( 44 )$&$0.42 ( 11 )$&$0.51$\\
$0.13970$&$1.72391 ( 57 )$&$0.1411085 ( 21 )$&$-1.417 ( 12 )$&$0.642 ( 43 )$&$0.43 ( 11 )$&$0.50$\\
$0.13980$&$1.72395 ( 56 )$&$0.1410623 ( 21 )$&$-1.418 ( 12 )$&$0.643 ( 43 )$&$0.43 ( 11 )$&$0.49$\\
$0.13990$&$1.72398 ( 56 )$&$0.1410159 ( 21 )$&$-1.420 ( 12 )$&$0.644 ( 42 )$&$0.43 ( 11 )$&$0.49$\\
$0.14000$&$1.72401 ( 56 )$&$0.1409692 ( 21 )$&$-1.421 ( 12 )$&$0.645 ( 42 )$&$0.43 ( 11 )$&$0.48$\\
$0.14010$&$1.72404 ( 55 )$&$0.1409220 ( 21 )$&$-1.422 ( 12 )$&$0.645 ( 42 )$&$0.44 ( 11 )$&$0.48$\\
$0.14020$&$1.72405 ( 55 )$&$0.1408747 ( 59 )$&$-1.423 ( 12 )$&$0.646 ( 42 )$&$0.44 ( 11 )$&$0.48$\\
$0.14030$&$1.72406 ( 55 )$&$0.1408273 ( 21 )$&$-1.425 ( 12 )$&$0.646 ( 42 )$&$0.44 ( 11 )$&$0.47$\\
$0.14040$&$1.72405 ( 54 )$&$0.1407798 ( 22 )$&$-1.426 ( 12 )$&$0.646 ( 41 )$&$0.44 ( 11 )$&$0.47$\\
$0.14050$&$1.72403 ( 53 )$&$0.1407324 ( 28 )$&$-1.428 ( 12 )$&$0.646 ( 41 )$&$0.44 ( 11 )$&$0.46$\\
$0.14060$&$1.72400 ( 53 )$&$0.1406849 ( 22 )$&$-1.429 ( 11 )$&$0.647 ( 40 )$&$0.44 ( 11 )$&$0.45$\\
$0.14070$&$1.72396 ( 52 )$&$0.1406374 ( 22 )$&$-1.431 ( 11 )$&$0.647 ( 39 )$&$0.45 ( 11 )$&$0.43$\\
$0.14080$&$1.72390 ( 51 )$&$0.1405899 ( 22 )$&$-1.433 ( 11 )$&$0.646 ( 39 )$&$0.45 ( 10 )$&$0.43$\\
$0.14090$&$1.72384 ( 50 )$&$0.1405421 ( 22 )$&$-1.434 ( 11 )$&$0.646 ( 38 )$&$0.45 ( 10 )$&$0.42$\\
$0.14100$&$1.72376 ( 49 )$&$0.1404941 ( 27 )$&$-1.436 ( 11 )$&$0.645 ( 38 )$&$0.44 ( 10 )$&$0.42$\\
$0.14110$&$1.72368 ( 49 )$&$0.1404459 ( 22 )$&$-1.437 ( 11 )$&$0.643 ( 38 )$&$0.44 ( 10 )$&$0.42$\\
$0.14120$&$1.72359 ( 49 )$&$0.1403976 ( 23 )$&$-1.439 ( 11 )$&$0.641 ( 38 )$&$0.44 ( 10 )$&$0.43$\\
$0.14130$&$1.72349 ( 49 )$&$0.1403490 ( 22 )$&$-1.440 ( 11 )$&$0.639 ( 38 )$&$0.43 ( 10 )$&$0.44$\\
$0.14140$&$1.72338 ( 49 )$&$0.1403004 ( 23 )$&$-1.441 ( 11 )$&$0.636 ( 38 )$&$0.43 ( 10 )$&$0.45$\\
$0.14150$&$1.72326 ( 50 )$&$0.1402515 ( 23 )$&$-1.442 ( 11 )$&$0.633 ( 38 )$&$0.42 ( 10 )$&$0.47$\\
$0.14160$&$1.72312 ( 50 )$&$0.1402023 ( 27 )$&$-1.443 ( 11 )$&$0.631 ( 38 )$&$0.42 ( 10 )$&$0.48$\\
$0.14170$&$1.72299 ( 49 )$&$0.1401529 ( 28 )$&$-1.443 ( 11 )$&$0.631 ( 38 )$&$0.42 ( 10 )$&$0.47$\\
\end{tabular} 
\end{ruledtabular} 
\end{table}

\begin{figure}[!hbt]
  \centering
  \includegraphics[bb= 0 0 340 340, width=0.5\textwidth]{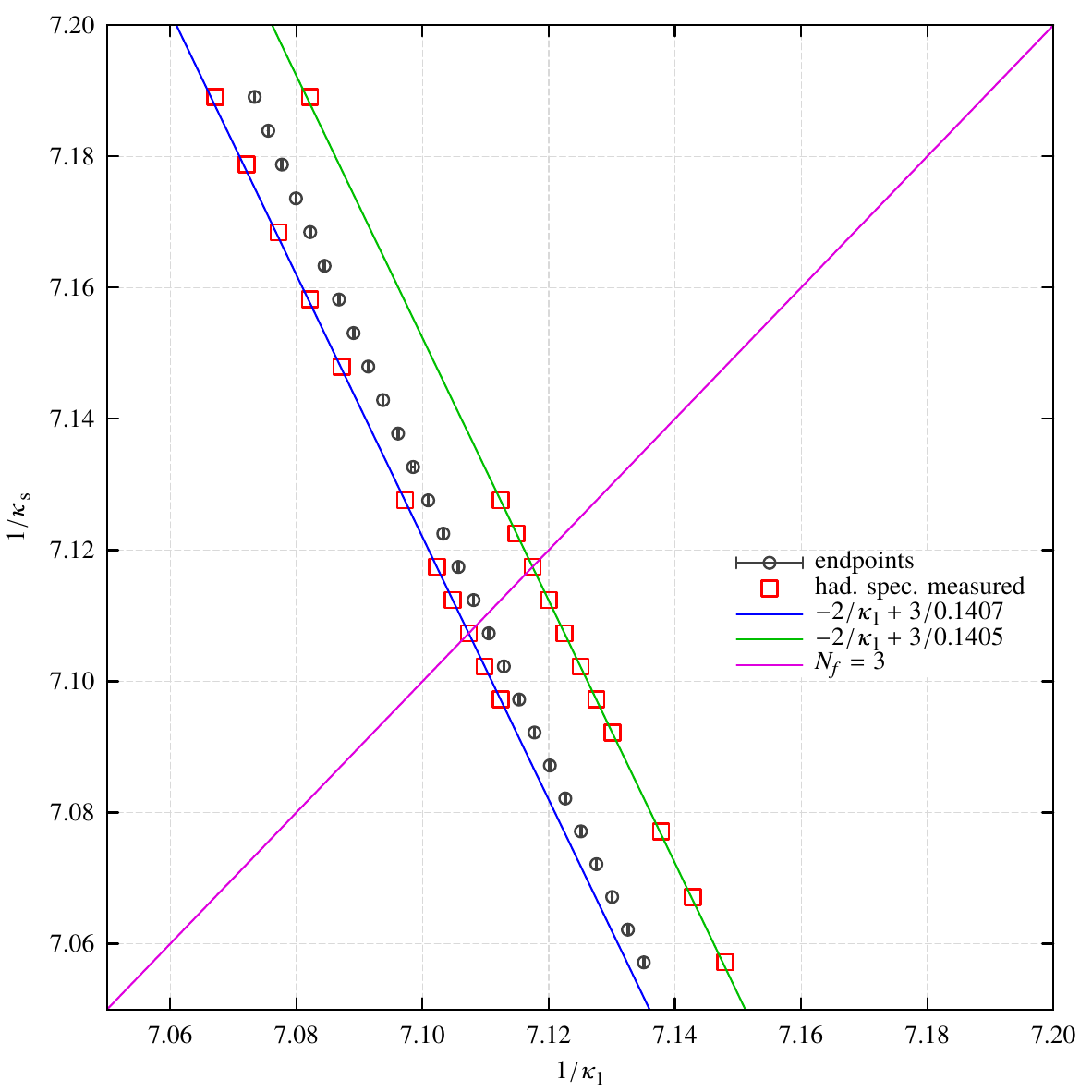}
  \caption{Results for the critical endpoints in the plane of  $1/\kappa_{\rm l}$ and $1/\kappa_{\rm s}$ (black open circles).  Also shown (red squares) are the points where zero temperature simulations are carried out to calculate hadron masses.  SU(3)-symmetric line is drawn in pink, while for green and blue lines the sum of three quark masses is constant.}
  \label{fig:klks}
\end{figure}

In Fig.~\ref{fig:klks}, we plot $1/\kappa_{\rm s, E}$ as a function of  $1/\kappa_{\rm l, E}$ by open circles; they represent our estimate of CEL.  We also plot the points where we have 
performed zero temperature simulations to calculate the pseudo-scalar meson masses. 
We have generated $O(500)$ configurations at $\beta=1.72$ on a $16^3 \times 32$ lattice for each $(\kappa_{\rm l}, \kappa_{\rm s})$.
The simulation parameters together with the Wilson flow scale $\sqrt{t_0}/a$~\cite{wilflow} and pseudo-scalar meson masses are summarized in Table~\ref{tab:PS}.

To calculate the pseudo scalar meson masses along CEL, a linear interpolation in Ward identity quark masses is sufficiently good in such a tiny parameter region.  
Thus, we perform a fit constrained by flavor symmetry of form, 
\begin{equation}
\begin{split}
(am_{\pi})^2 		=& 2 A_1 		\bar m_l +  2 A_2 		\bar m_s \,,\\
(am_{K})^2 		=& (A_1  +A_3) 	\bar m_l +  (A_2+A_4) 	\bar m_s \,,\\
(am_{\eta_s})^2 	=& 2  A_3 		\bar m_l +  2A_4 		\bar m_s \,,\\
\sqrt{t_0}/a 		=& A_5 + 2 A_6 	\bar m_l +  A_6 		\bar m_s \,,\\
\end{split}
\end{equation}
with
$\bar m_l 	= 1/\kappa_{\rm l} 	- 1/\kappa_0$, 
$\bar m_s 	= 1/\kappa_{\rm s} 	- 1/\kappa_0$.
We obtain
$\kappa_0=  0.1419248(25)$,
$A_1=  3.0896(55)$,
$A_2=  0.8494(37)$,
$A_3=  1.8091(54)$,
$A_4=  2.1228(38)$,
$A_5=  0.87742(43)$,
$A_6=  -0.5034(20)$,
and $\chi^2/\dof = 16.9$.

\begin{table}[!hbt]
\caption{$\sqrt{t_0}/a$ and pseudo-scalar masses in lattice unit at $\beta=1.72$ on $16^3 \times32$}
\label{tab:PS}%
\begin{ruledtabular}
\begin{tabular}{ll|cccc}
$\kappa_{\rm l}$ & $\kappa_{\rm s}$ & $\sqrt{t_0}/a$ & $ am_{\pi} $ &  $am_{K}$ &  $am_{\eta_s}$ \\
\colrule
  $0.140600$ &   $0.140900$ &     $0.78338(31)$ &      $0.7054(10)$ &      $0.6917(10)$ &      $0.6778(10)$\\
  $0.140650$ &   $0.140800$ &     $0.78324(33)$ &      $0.7015(10)$ &      $0.6947(10)$ &      $0.6878(10)$\\
  $0.140700$ &   $0.140700$ &     $0.78391(32)$ &      $0.6960(11)$ &      $0.6960(11)$ &      $0.6960(11)$\\
  $0.140750$ &   $0.140600$ &     $0.78369(31)$ &      $0.6925(11)$ &      $0.6994(11)$ &      $0.7063(11)$\\
  $0.140800$ &   $0.140500$ &     $0.78331(31)$ &      $0.6881(13)$ &      $0.7019(13)$ &      $0.7154(13)$\\
  $0.140900$ &   $0.140300$ &     $0.78397(43)$ &      $0.6755(13)$ &      $0.7032(13)$ &      $0.7300(13)$\\
  $0.141100$ &   $0.139900$ &     $0.78420(39)$ &      $0.6573(12)$ &      $0.7133(12)$ &      $0.7654(12)$\\
  $0.141200$ &   $0.139700$ &     $0.78538(33)$ &      $0.6444(10)$ &      $0.7146(10)$ &      $0.7788(09)$\\
  $0.141300$ &   $0.139500$ &     $0.78639(25)$ &      $0.6319(12)$ &      $0.7170(11)$ &      $0.7936(11)$\\
  $0.141400$ &   $0.139300$ &     $0.78784(27)$ &      $0.6174(12)$ &      $0.7174(11)$ &      $0.8059(10)$\\
  $0.141500$ &   $0.139100$ &     $0.78799(45)$ &      $0.6018(14)$ &      $0.7173(12)$ &      $0.8176(11)$\\
\colrule
  $0.139900$ &   $0.141700$ &     $0.77075(28)$ &      $0.7996(11)$ &      $0.7226(11)$ &      $0.6368(12)$\\
  $0.140000$ &   $0.141500$ &     $0.77037(26)$ &      $0.7895(16)$ &      $0.7252(17)$ &      $0.6549(18)$\\
  $0.140100$ &   $0.141300$ &     $0.77002(25)$ &      $0.7835(09)$ &      $0.7321(09)$ &      $0.6770(10)$\\
  $0.140250$ &   $0.141000$ &     $0.76907(26)$ &      $0.7722(09)$ &      $0.7401(09)$ &      $0.7067(09)$\\
  $0.140300$ &   $0.140900$ &     $0.76882(27)$ &      $0.7696(11)$ &      $0.7438(11)$ &      $0.7172(11)$\\
  $0.140350$ &   $0.140800$ &     $0.76894(27)$ &      $0.7652(11)$ &      $0.7459(11)$ &      $0.7260(11)$\\
  $0.140400$ &   $0.140700$ &     $0.76868(27)$ &      $0.7620(14)$ &      $0.7491(14)$ &      $0.7360(14)$\\
  $0.140450$ &   $0.140600$ &     $0.76857(27)$ &      $0.7583(11)$ &      $0.7519(11)$ &      $0.7453(11)$\\
  $0.140500$ &   $0.140500$ &     $0.76843(27)$ &      $0.7542(12)$ &      $0.7542(12)$ &      $0.7542(12)$\\
  $0.140550$ &   $0.140400$ &     $0.76875(27)$ &      $0.7491(10)$ &      $0.7555(10)$ &      $0.7620(10)$\\
  $0.140600$ &   $0.140300$ &     $0.76887(28)$ &      $0.7428(11)$ &      $0.7558(11)$ &      $0.7687(11)$\\
  $0.141200$ &   $0.139100$ &     $0.77111(26)$ &      $0.6845(11)$ &      $0.7780(10)$ &      $0.8619(10)$\\
\end{tabular}
\end{ruledtabular}
\end{table}

\begin{figure}[!hbt]
  \centering
  \includegraphics[bb= 0 0 340 340, width=0.5\textwidth]{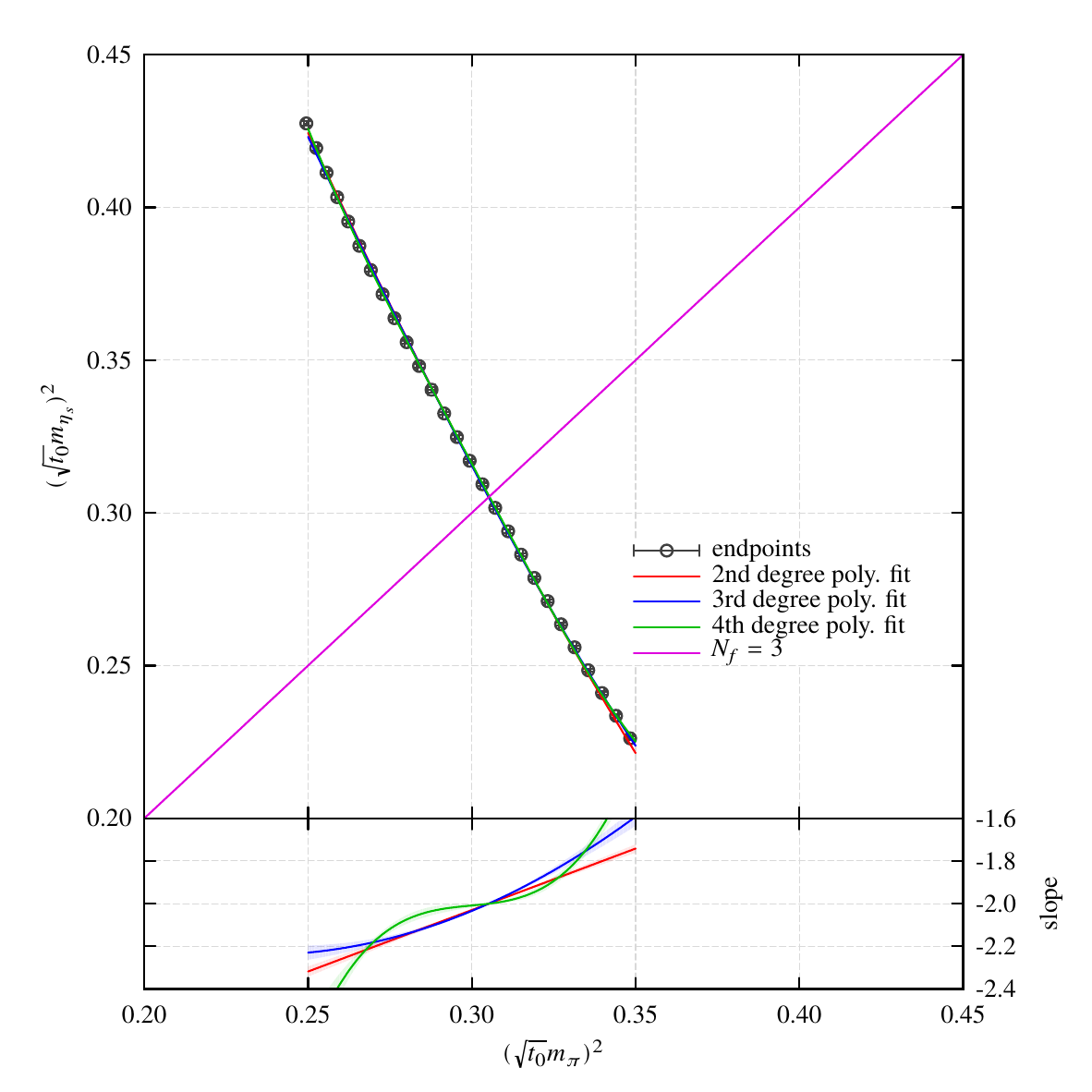}
  \caption{
  CEL in the plane of $(\sqrt{t_0}m_{\pi})^2$ and $(\sqrt{t_0}m_{\eta_s})^2$ and the slope along CEL calculated by the fit in eq. (\ref{eq:fit}).  Pink line denotes the line of SU(3) symmetry ( $N_{\rm f}=3$). Three types of polynomial fitting results are overlapping in upper plot.}
  \label{fig:mpsE}
\end{figure}

Figure \ref{fig:mpsE} shows the results for $(\sqrt{t_0}m_\pi)^2$ and $(\sqrt{t_0}m_{\eta_s})^2$ along  CEL.  
We fit the data points by the following three polynomial functions up to fourth order in order to check higher order contributions against the second derivative.
\begin{equation}
\label{eq:fit}
\begin{split}
f_2(x)=&b_{2,0}-2(x-b_{2,0})+b_{2,1}(x-b_{2,0})^2\,, \\
f_3(x)=&b_{3,0}-2(x-b_{3,0})+b_{3,1}(x-b_{3,0})^2+b_{3,2}(x-b_{3,0})^3\,,\\
f_4(x)=&b_{4,0}-2(x-b_{4,0})+b_{4,1}(x-b_{4,0})^2+b_{4,2}(x-b_{4,0})^3+b_{4,3}(x-b_{4,0})^4\,,\\
\end{split}
\end{equation}
where $f(x)=(\sqrt{t_0}m_{\eta_s})^2$ and $x=(\sqrt{t_0}m_{\pi})^2$.  
The fit results are given in Table~\ref{tab:fit}.
We find that the results are reasonably consistent with a slope of  $-2$ and a positive second derivative at CEP. 

\begin{table}[!hbt]
\caption{Results of polynomial fitting for CEL.}
\label{tab:fit}%
\begin{ruledtabular}
\begin{tabular}{l|ccccc}
func. & $b_0$ &  $ b_1 $ &  $b_2$ & $b_3$ & $\chi^2/\dof$ \\
\colrule
$ f_2 $ & $0.305194(83)$ & $    2.88(22)$ & $           -$ & $           -$ & $  1.619$ \\
$ f_3 $ & $0.305128(84)$ & $    3.44(26)$ & $   16.4(4.3)$ & $           -$ & $  1.039$ \\
$ f_4 $ & $ 0.30534(11)$ & $    0.96(71)$ & $   27.2(5.3)$ & $   1212(325)$ & $  0.481$ \\
\end{tabular}
\end{ruledtabular}
\end{table}

\section{Summary}\label{sec:sum}

We have investigated CEL of the finite temperature phase transition of  QCD 
with non-perturbatively $O(a)$-improved Wilson-clover fermion action 
around the SU(3)-symmetric point at zero chemical potential and $N_{\rm T}=6$.
Our method of  kurtosis intersection point analysis aided by multi-parameter, multi-ensemble reweighting works well.
As results, we could precisely determine CEL over a range $0.25 \le (\sqrt{t_0}m_{\pi})^2 \le 0.35$ sandwitching the SU(3)-symmetric point at $(\sqrt{t_0}m_{\pi})^2\approx 0.3$, and 
found $-2$ for the slope and a positive second derivative around $m^{\rm sym}$
on the Columbia plot.

We need to add to remarks on our results. 
First, for zero temperature simulations, we used a slightly different $\beta$ than $\beta_\rmE$. 
We think that this difference will not change our conclusion since it is only an effect of 0.3\% or so in hadron mass values. 
Second, our study is conducted at just single lattice spacing of $a \approx 0.19$ fm.  
We are pursuing simulations with a larger $N_{\rm T}$ to obtain conclusive results, especially for the second derivative. 
Finally, the physical point at $(\sqrt{t_0}m_{\pi})^2\approx 0.01$ is quite far from the SU(3)-symmetric point.  Thus, $N_{\rm f}=2+1$ simulations are needed to investigate CEL as it approaches the physical point. 

\begin{acknowledgments}
The BQCD code~\cite{BQCD} was used in this work.
This research used computational resources of 
HA-PACS and COMA provided by Interdisciplinary Computational Science Program in Center for Computational Sciences, University of Tsukuba,
System E at Kyoto University through the HPCI System Research project (Project ID:hp140180) and
PRIMERGY CX400 tatara at Kyushu University.
This work is supported by JSPS KAKENHI Grant Numbers 23740177 and 26800130,
FOCUS Establishing Supercomputing Center of Excellence and
Kanazawa University SAKIGAKE Project.
\end{acknowledgments}

\end{document}